\documentclass[prd,aps,a4paper,nofootinbib,notitlepage,twocolumn]{revtex4-1}
\usepackage[a4paper, hdivide={1.91cm,,1.165cm}, vdivide={1.83cm,,4.1cm}]{geometry}

\usepackage{amsmath,amssymb}
\usepackage{graphicx,multirow}
\usepackage{units}
\usepackage{supertabular}
\usepackage[hyperfootnotes=false]{hyperref}

\newcommand{\dd}{\ensuremath{\text{d}}}

\newcommand{\Element}[3]{${}^{#3}_{#2}\text{#1}$}

\allowdisplaybreaks

\begin{document}

\title{Isotope dependence of muon decay in orbit}

\author{Julian Heeck}
\email{heeck@virginia.edu}
\affiliation{Department of Physics, University of Virginia,
Charlottesville, Virginia 22904-4714, USA}

\author{Robert Szafron}
\email{rszafron@bnl.gov}
\affiliation{Department of Physics, Brookhaven National Laboratory, Upton, New York, 11973, USA}

\author{Yuichi Uesaka}
\email{uesaka@ip.kyusan-u.ac.jp}
\affiliation{Faculty of Science and Engineering, Kyushu Sangyo University,
2-3-1 Matsukadai, Higashi-ku, Fukuoka 813-8503, Japan}

\begin{abstract}
The decay $\mu^-\to e^- \bar{\nu}_e\nu_\mu$ of a muon that is bound to a nucleus poses an unavoidable background for experiments such as Mu2e, COMET, and DeeMe searching for lepton-flavor-violating $\mu^-\to e^-$ conversion and thus requires a precise understanding. We calculate the electron-energy spectra of muon decays in orbit near the electron energy endpoint for all stable isotopes and estimate uncertainties due to the nuclear charge distribution. Our results enable background studies of mixed or enriched target materials that are necessary for the next generation of $\mu^-\to e^-$ conversion experiments.
\end{abstract}

\maketitle


\section{Introduction}

The search for lepton flavor violation (LFV) is a sensitive probe of physics beyond the Standard Model~\cite{Bernstein:2013hba}. Rare \emph{muon} decays are a particularly interesting avenue due to their long lifetime and the relative ease to produce them in large numbers. Particularly clean processes are the decays $\mu \to e \gamma$ and $\mu\to 3 e$ as well as the coherent $\mu^-\to e^-$ conversion in the field of a nucleus, $\mu^- +(A,Z) \to e^- +(A,Z)$, the latter being especially sensitive to effective LFV operators involving quarks~\cite{Crivellin:2017rmk,Davidson:2018kud}.
An immense jump in sensitivity to $\mu^-\to e^-$ conversion in nuclei is expected in the near future with experiments such as DeeMe~\cite{Natori:2014yba,Teshima:2019orf}, COMET~\cite{Adamov:2018vin,Moritsu:2021fns}, and Mu2e~\cite{Bartoszek:2014mya,Yucel:2021vir}. 
Since the $\mu^-\to e^-$ conversion rate depends on the target nucleus (notably its charge $Z$), it is possible to differentiate between different LFV effective operators using complementary target nuclei in the event of a positive signal observation~\cite{Kitano:2002mt,Cirigliano:2009bz,Bartolotta:2017mff,Davidson:2017nrp,Davidson:2018kud,Davidson:2020hkf}.
DeeMe will use graphite or silicon carbide targets while COMET and Mu2e study $\mu^-\to e^-$ conversion in aluminium. Previous experiments set limits using copper~\cite{Bryman:1972rf}, sulfur~\cite{Badertscher:1980bt}, lead~\cite{SINDRUMII:1996fti}, titanium~\cite{SINDRUMII:1993gxf}, and gold~\cite{SINDRUMII:2006dvw} targets.

$\mu^-\to e^-$ conversion of bound muons produces approximately monoenergetic electrons with energy 
\begin{align}
E_\text{end} =  m_\mu - E_b - E_\text{recoil} \,,
\label{eq:endpoint}
\end{align}
where $E_b\simeq \alpha^2 Z^2 m_\mu/2$ and $E_\text{recoil}= ( m_\mu - E_b)^2/(2 m_N)$ are small corrections due to the muon's binding and nuclear recoil, respectively, in a nucleus with charge $Z$ and mass $m_N$.
While electron energies $\sim m_\mu$ are seemingly far away from the typical electron energies of the competing $\mu\to e \nu\nu$ decay in orbit (DIO), the presence of a heavy nucleus ensures that this Standard-Model decay distribution has a tail up to $E_\text{end}$, providing an irreducible background. Precise calculations are necessary to predict and understand this background DIO spectrum near the electron-energy endpoint in order to chose the optimal signal window.
Such calculations exist~\cite{Czarnecki:1998iz,Czarnecki:2011mx,Czarnecki:2011ei,Szafron:2015kja,Szafron:2015wbm,Szafron:2016cbv}, most importantly for aluminium, but are lacking sufficient precision for other target nuclei.

Here we set out to provide the DIO spectrum near the electron endpoint -- as relevant for $\mu^-\to e^-$ conversion experiments -- for all stable nuclei, including isotopes.
We use the most up-to-date data for the necessary nuclear charge distributions and include the dominant radiative corrections due to the soft photon bremsstrahlung.

\section{Muon decay in orbit}

Calculating the $\mu\to e \nu\nu$ decay of a bound muon requires solving Dirac equation for both the bound muon and the outgoing electron in the electric field of the nucleus~\cite{Porter:1951zz,Haenggi:1974hp,Shanker:1981mi,Watanabe:1987su,Watanabe:1993emp,Shanker:1996rz}. For spherically symmetric charge distribution $\rho$, the electric potential $V$ is given by
\begin{align}
V (r) = - e \int_0^\infty\dd \tilde{r}\, \tilde{r}^2 \left[\frac{\theta (r-\tilde{r}) }{r}+\frac{\theta (\tilde{r}-r) }{\tilde{r}}\right] \rho (\tilde{r}) \,,
\label{eq:potential}
\end{align}
with the Heaviside $\theta$ function.
The charge distribution is normalized as follows:
\begin{align}
4\pi \int_0^\infty\dd r\, r^2 \rho(r) = Z e\,.
\label{eq:normalization}
\end{align}
For a given charge distribution the potential $V$ can be calculated and the Dirac equation for muon and electron be solved numerically. The relevant formulae are given in  Ref.~\cite{Czarnecki:2011mx}, which we follow closely. The Dirac equation provides us with the muon binding energy $E_b$ and the wavefunctions of bound muon and free electron in the Coulomb background, which ultimately determine the decay rate.
Ref.~\cite{Czarnecki:2011mx} also provided an expansion of the DIO spectrum near the endpoint of the form $\dd \Gamma/\dd E_e \simeq \Gamma_0 B (E_e - E_\text{end})^5$, where $\Gamma_0 = G_F^2 m_\mu^5/(192\pi^3)$ is the leading-order free-muon decay rate and $B$ is a coefficient of mass dimension $-6$ that depends on several overlap integrals over products of wavefunctions.

Here, we improve the endpoint expansion of Ref.~\cite{Czarnecki:2011mx} by implementing the leading radiative corrections:
\begin{align}
\frac{1}{\Gamma_0} \frac{\dd \Gamma }{\dd E_e} \Big|_{E_e \sim E_\text{end}'} =  B \, E_\text{end}'^5 \left( 1 - \frac{E_e}{{E_\text{end}'}}\right)^{5 + \delta} ,
\label{eq:spectrum}
\end{align}
with $\delta = \alpha (2\log [2m_\mu/m_e] -2)/\pi\simeq 0.023$~\cite{Szafron:2015kja} coming from soft photon radiation and a shifted endpoint energy~\cite{Szafron:2016cbv}
\begin{align}
E_\text{end}' \equiv E_\text{end}  +\frac{\alpha  m_\mu \left( Z \alpha\right)^2 }{\pi} \left( \frac{11}{9} - \frac{2}{3} \log \left[ \frac{2 m_\mu Z \alpha}{m_e}\right]\right) 
\label{eq:endpoint_shift}
\end{align}
due to the vacuum polarization effects. Here, $E_\text{end}$ is given by Eq.~\eqref{eq:endpoint}.
Notice that $E_\text{end}'$ coincides with the electron energy in $\mu^-\to e^-$ conversion. In the above formulas, $\alpha = 7.297\,352\, 5693(11)\times 10^{-3}$ \cite{Tiesinga:2021myr} is the fine structure constant. 

For a target that consists of several isotopes, the above spectrum $\dd \Gamma /\dd E_e$ is to be summed over the isotopes, weighted by their abundance. This is necessary because both the coefficient $B$ and the binding energy $E_b$ depend to a small degree on the number of neutrons in the nucleus. In the following we will determine how large this dependence is.

\section{Nuclear charge distributions}

The electric-charge distribution of nuclei will be approximated as spherically symmetric in the following and denoted as $\rho (r)$, normalized according to Eq.~\eqref{eq:normalization}.
Experimental information about $\rho$ can be obtained via spectroscopy in (muonic) atoms and through scattering. For example, electron--nucleus scattering cross sections with momentum transfer $q$ in the Born approximation are proportional to the squared form factor $|F(q)|^2$, which is related to $\rho(r)$ by Fourier transformation:
\begin{align}
F(q) &= \frac{4\pi}{Z e}\int_0^\infty \rho(r)\frac{\sin (q r)}{q r} r^2 \dd r\\
\Leftrightarrow \quad\rho (r) &= \frac{4\pi Z e}{(2\pi)^3}\int_0^\infty F(q)\frac{\sin (q r)}{q r} q^2 \dd q \,.
\end{align}
We have normalized $F(q)$ so that it gives $1$ for a point charge.
Since scattering experiments can only be performed over a finite range of $q$, it is not possible to obtain $\rho$ model-independently; rather, once a parametrization for $\rho (r)$ is chosen, $F(q)$ can be calculated and fitted to data.

Popular parametrizations for spherically symmetric charge distributions with varying degrees of complexity are listed below:
\begin{enumerate}
\item Three-parameter Fermi model (3pF)~\cite{Yennie:1954zz,Hahn:1956zz}:
\begin{align}
\rho (r) = \frac{\rho_0 }{1 + \exp\left(\frac{r-c}{z}\right)}\left(1 + w\frac{ r^2}{c^2}\right) .
\end{align}
The two-parameter Fermi model (2pF) can be obtained as the special case $w=0$; the one-parameter Fermi function (1pF) is defined here through $w=0$ and $z = \unit[0.52]{fm}$ (which corresponds to a constant surface thickness of $\unit[2.3]{fm}$). 
\item Three-parameter Gaussian model (3pG)~\cite{Hahn:1956zz}:
\begin{align}
\rho (r) = \frac{\rho_0 }{1 + \exp\left(\frac{r^2-c^2}{z^2}\right)}\left(1 + w\frac{ r^2}{c^2}\right) .
\end{align}
\item Modified-harmonic oscillator model (MHO)~\cite{Hofstadter:1957wk}:
\begin{align}
\rho (r) = \rho_0 \left(1 + w\frac{ r^2}{a^2}\right) \text{e}^{-r^2/a^2} .
\end{align}
\item Fourier--Bessel expansion (FB)~\cite{Dreher:1974pqw}:
\begin{align}
\rho (r) = \begin{cases}
\sum_k^n a_k j_0 \left(\frac{k \pi r}{R}\right) , & \, \, r\leq R\,,\\
0\,, & \,\, r>R\,.
\end{cases}
\end{align}
Here, $a_{1,\dots,n}$ are the FB coefficients and $j_0(x) = \sin(x)/x$ the spherical Bessel function of order zero.
$n\leq 18$ for our data.
\item Sum of Gaussians (SOG)~\cite{Sick:1974suq}:
\begin{align}
\rho (r) = Z e \sum_k^n Q_k\,\frac{ \text{e}^{-\frac{(r-R_k)^2}{\gamma^2}}+\text{e}^{-\frac{(r+R_k)^2}{\gamma^2}}}{2\pi^{3/2} \gamma^3 (1+ 2 R_k^2/\gamma^2)}\,,
\end{align}
with $\sum_k^n Q_k = 1$.
\end{enumerate}
The normalization $\rho_0$ can be calculated analytically in all cases and the potential $V$ in Eq.~\eqref{eq:potential} as well, except for the 3pG model, and are given in appendix~\ref{app:potentials}.

The simplest parametrization, 1pF, takes as input only the nuclear charge radius, listed for all elements and isotopes of interest in Ref.~\cite{Angeli:2013epw}. For many isotopes this is the only available parametrization, making it extremely valuable despite its lack of substructure.
For the multi-parameter parametrizations we use the available data tables from Refs.~\cite{DeVries:1987atn,Boeglin:1988yfc,Fricke:1995zz,Wesseling:1997zz,Kabir:2015igz}, generally choosing the newest possible data set for a given isotope.

For several parametrizations the errors on the fit parameters are either not given in the literature (e.g.~for FB) or not representative of the uncertainties in the charge distribution (e.g.~the errors on the charge radius relevant for 1pF are tiny).
To obtain an estimate for the error on our DIO spectrum from the uncertainty of the nuclear charge distribution we will compare the results of several parametrizations for a given isotope.

We restrict our survey to the 237 stable isotopes with natural abundance above $1\%$. Only half of these isotopes have a charge parametrization with two or more parameters, for the other half we have to rely on the simplistic 1pF model.

\section{Results}

For a given charge distribution we solve the Dirac equations numerically using a fourth-order Runge--Kutta method to extract the muon binding energy $E_b$ and evaluate the coefficient $B$, which can then be put into Eq.~\eqref{eq:spectrum}.  Our results agree with Ref.~\cite{Czarnecki:2011mx} for the six elements presented there, using their charge parametrizations. Our results for the binding energy also match the point-charge approximation $E_b\simeq m_\mu(1-\sqrt{1- Z^2\alpha^2})$ for small/light nuclei~\cite{Gordon:1928,Darwin:1928}.
Table~\ref{tab:results} provides our results and will be discussed in the following.

\subsection{Binding energy}

\begin{figure}
\includegraphics[width=0.48\textwidth]{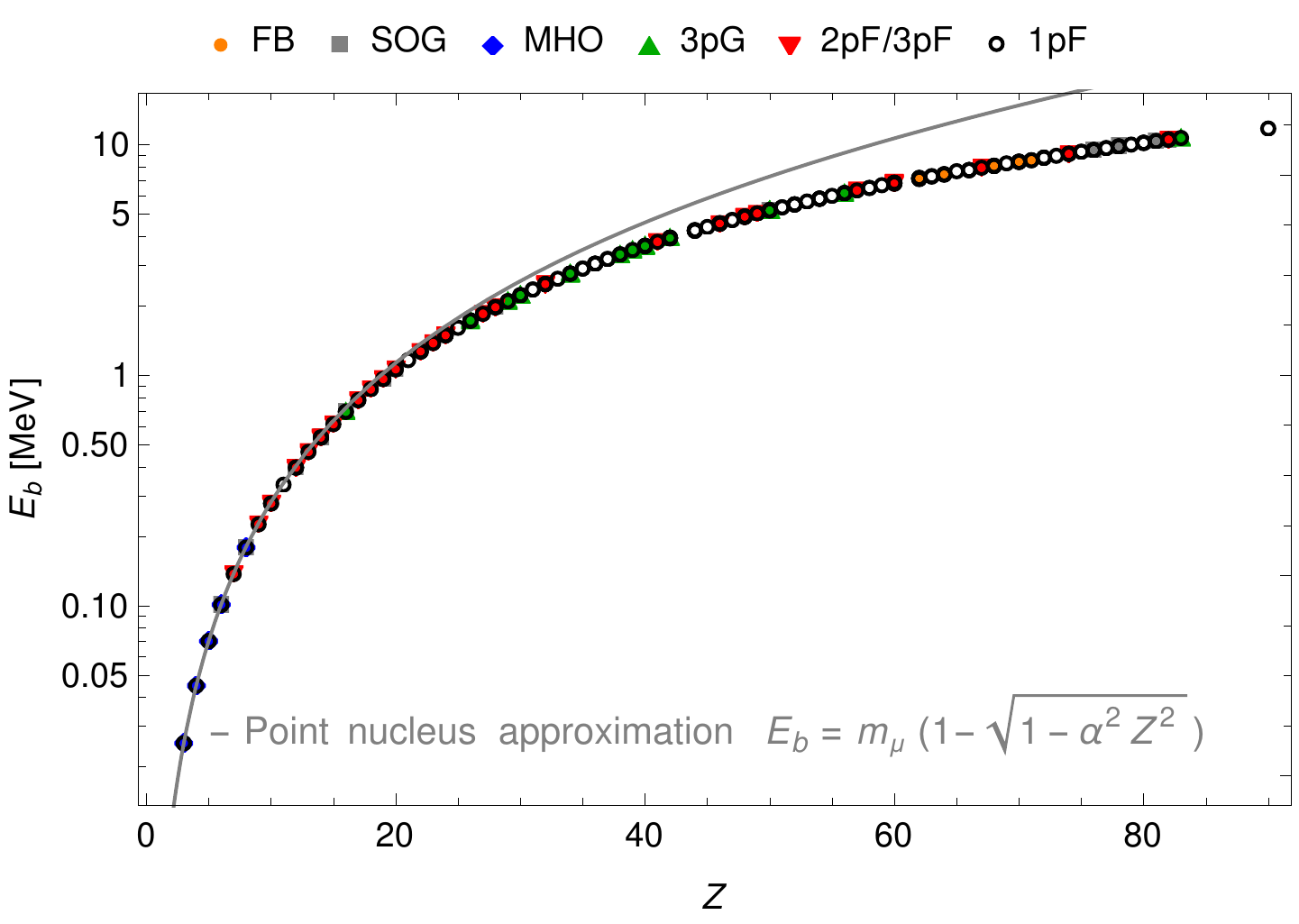}\vspace{4ex}
\includegraphics[width=0.48\textwidth]{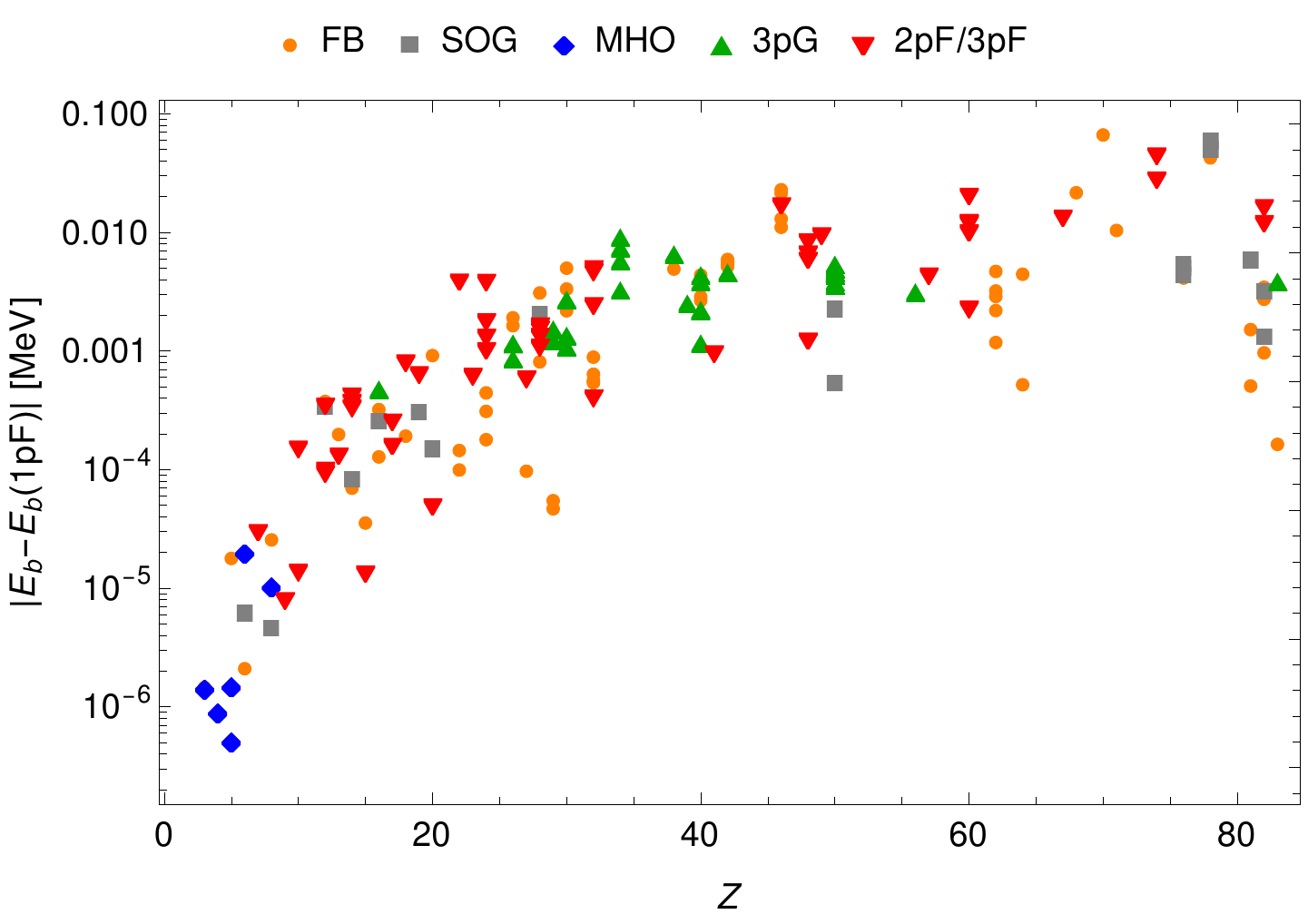}\vspace{4ex}
\includegraphics[width=0.48\textwidth]{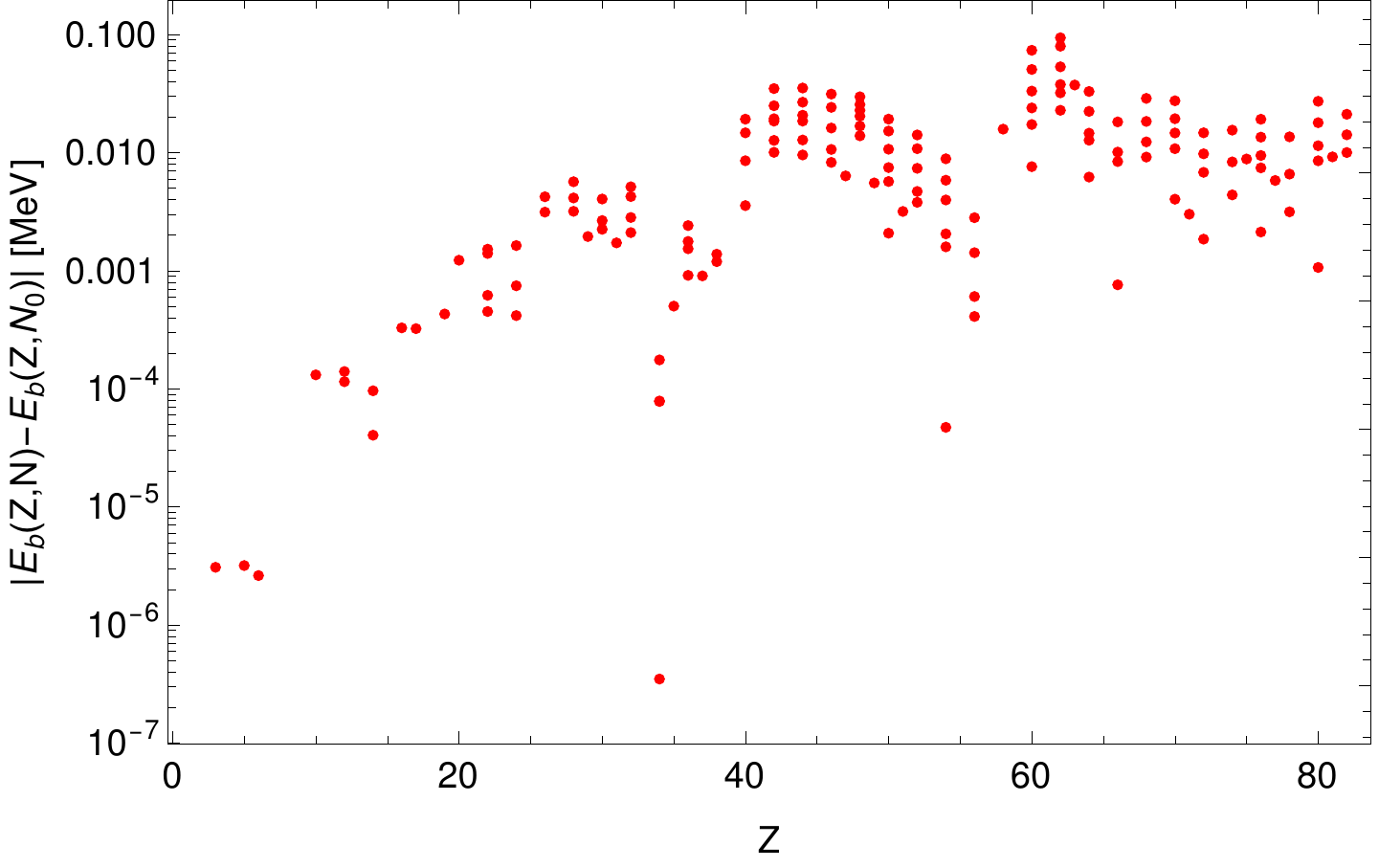}
\caption{
Top: muon binding energy $E_b$ vs $Z$ for all available paramtrizations.
Middle: difference of binding energies calculated with the 1pF charge distribution and other distributions.
Bottom: isotope dependence calculated using 1pF distributions, i.e.~the difference in binding energies between isotopes $(Z,N)$ and $(Z,N_0)$, with $N_0$ the smallest stable neutron number.
\label{fig:binding}}
\end{figure}

Our results for the muon binding energy $E_b$ as a function of $Z$ are shown in Fig.~\ref{fig:binding} (top) for all elements, isotopes, and available parametrizations. $E_b (Z)$ is very well behaved and agrees with the point-nucleus approximation $E_b= m_\mu(1-\sqrt{1- Z^2\alpha^2})$ for $Z< 20$. For $Z> 50$, the $Z$ dependence becomes linear and is approximately given by $E_b/(\alpha m_\mu) = -3.93 + 0.21 Z$.

Since the differences between the different parametrizations cannot be seen in Fig.~\ref{fig:binding} (top), we show in Fig.~\ref{fig:binding} (middle) the difference between the 1pF parametrization result with all the other available parametrizations. The relative difference is below $1\%$ for all $Z$ and markedly smaller for small $Z$. We take this as a sign that the endpoint energy does not depend strongly on the details of the charge distribution and is hence given by the 1pF results for all isotopes with at least $1\%$ accuracy, and even far better for smaller $Z$ (say $Z<20$).
Using the 1pF results we can illustrate the isotope dependence of $E_b$ by taking the difference between $E_b(Z,N)$ and $E_b(Z,N_0)$, $N_0$ being the neutron number of the lightest stable isotope. This is shown in Fig.~\ref{fig:binding} (bottom). The relative changes between isotopes are below $1\%$, once again smaller for smaller $Z$. Overall they are of similar size as the parametrization uncertainty.

\begin{figure}[t]
\includegraphics[width=0.48\textwidth]{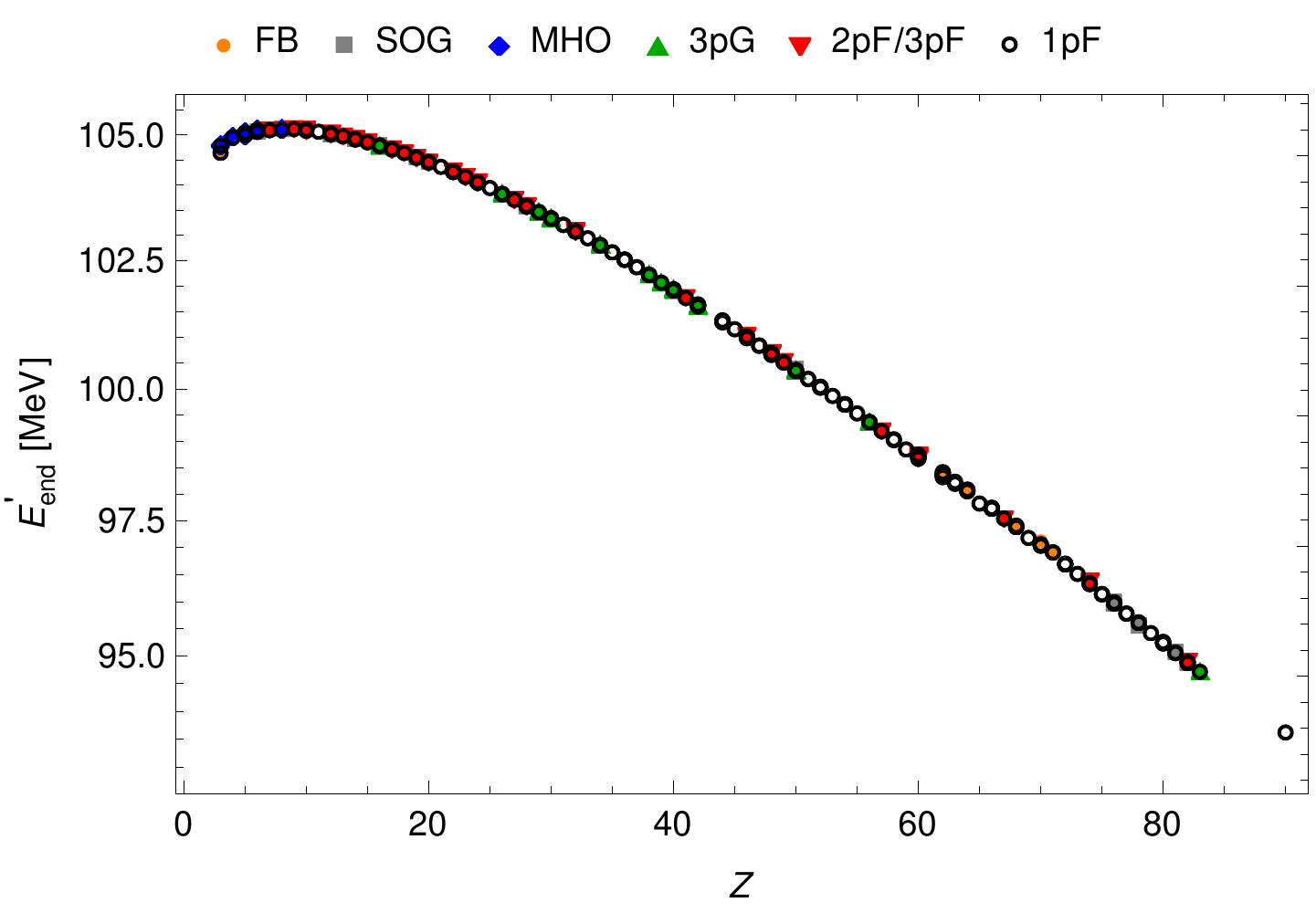}
\caption{
Endpoint energy $E_\text{end}'$ from Eq.~\eqref{eq:endpoint_shift} vs $Z$ for all isotopes in all available charge-distribution parametrizations. The relative uncertainty as well as isotope dependence is below one per mille.
}
\label{fig:endpoint}
\end{figure}

Ultimately, the quantity of interest for $\mu^-\to e^-$ conversion experiments is not the binding energy $E_b$ but the related endpoint energy $E_\text{end}'$, given by Eq.~\eqref{eq:endpoint_shift}. Since $E_b$ is itself only a small correction to the leading-order estimate $E_\text{end}'\sim m_\mu$, small uncertainties in $E_b$ are even less relevant in  $E_\text{end}'$. In fact, the relative parametrization uncertainty as well as the isotope dependence of  $E_\text{end}'$ are below one per mille for all $Z$.
Therefore, the endpoint energies (shown in Fig.~\ref{fig:endpoint}) are well approximated by the 1pF results, available for all isotopes of interest.

Notice that we have approximated the nuclear-recoil energy in Eq.~\eqref{eq:endpoint} at leading order in $1/m_N$, which is an excellent approximation for heavy nuclei; for light nuclei, e.g.~\Element{Li}{3}{6}, the subleading recoil terms can be of order $m_\mu^3/m_N^2\sim \unit[0.04]{MeV}$ and pose the largest remaining uncertainty of the endpoint energy.

\subsection{\texorpdfstring{$B$}{B} coefficient}

\begin{figure}[t]
\includegraphics[width=0.48\textwidth]{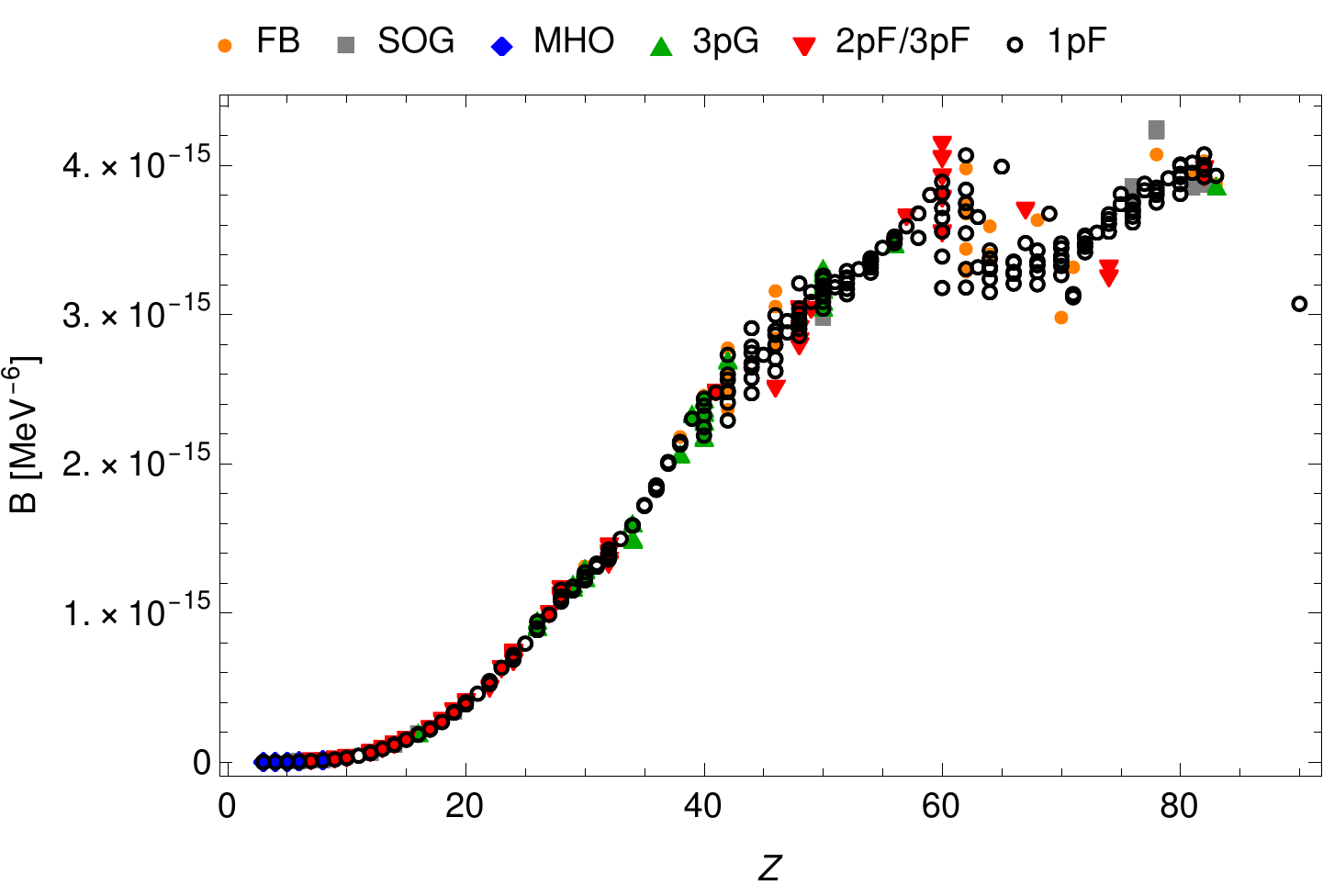}\vspace{3ex}
\includegraphics[width=0.48\textwidth]{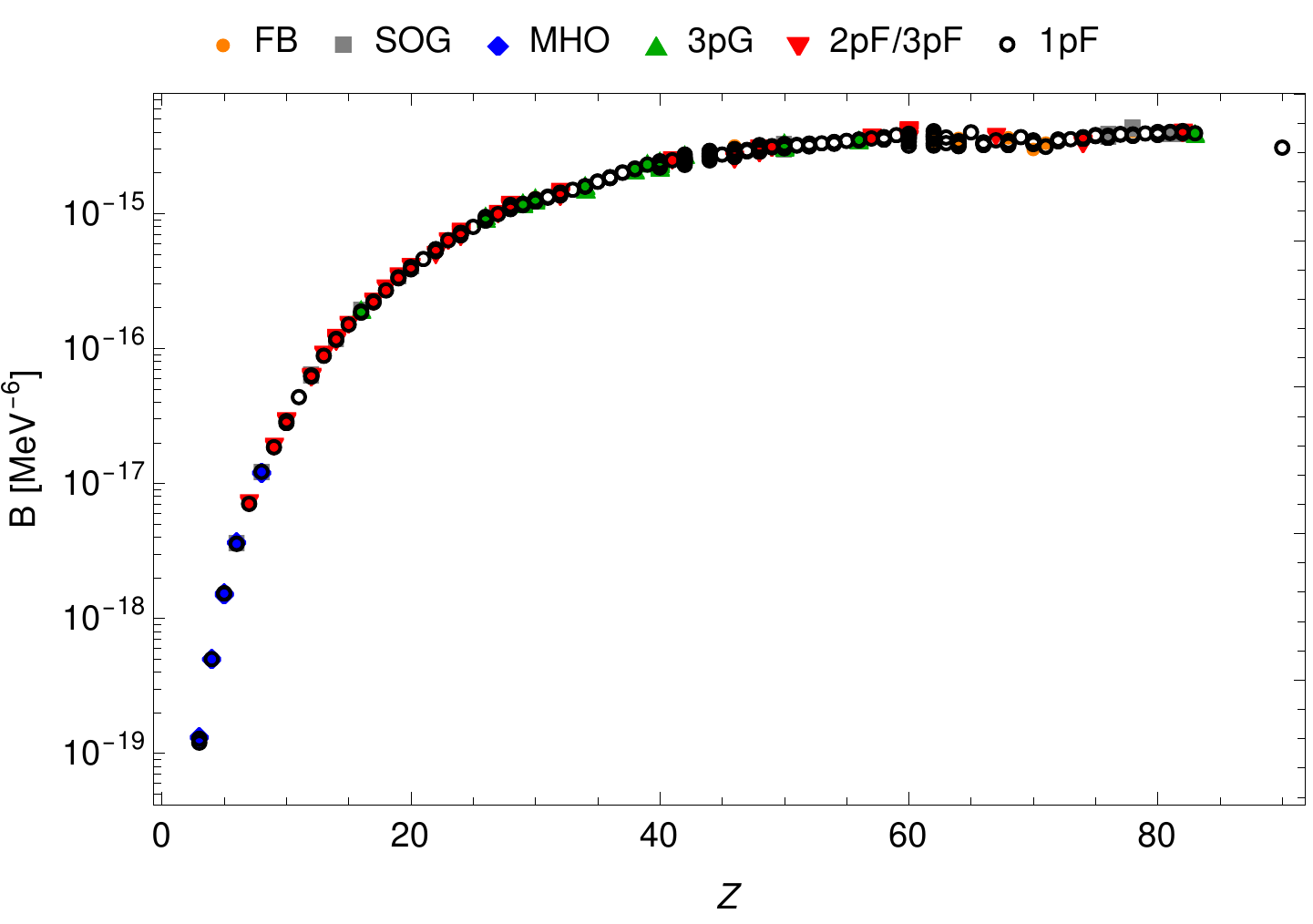}
\caption{
$B$ coefficient from Eq.~\eqref{eq:spectrum} on a linear scale (top) and on a log scale (bottom).
}
\label{fig:Bcoefficient}
\end{figure}

Having determined the electron endpoint for all isotopes, we can move on to the more difficult task of determining the $B$ coefficient in Eq.~\eqref{eq:spectrum}, which provides the normalization of the spectrum and hence the number of electrons within the signal window of a given $\mu^-\to e^-$ conversion experiment.
Our results are shown in Fig.~\ref{fig:Bcoefficient}, which illustrates that the dependence of $B$ on the charge distribution, and by extension the isotope, is much larger than for the binding or endpoint energies, at least for large $Z$.

\begin{figure}[t]
\includegraphics[width=0.48\textwidth]{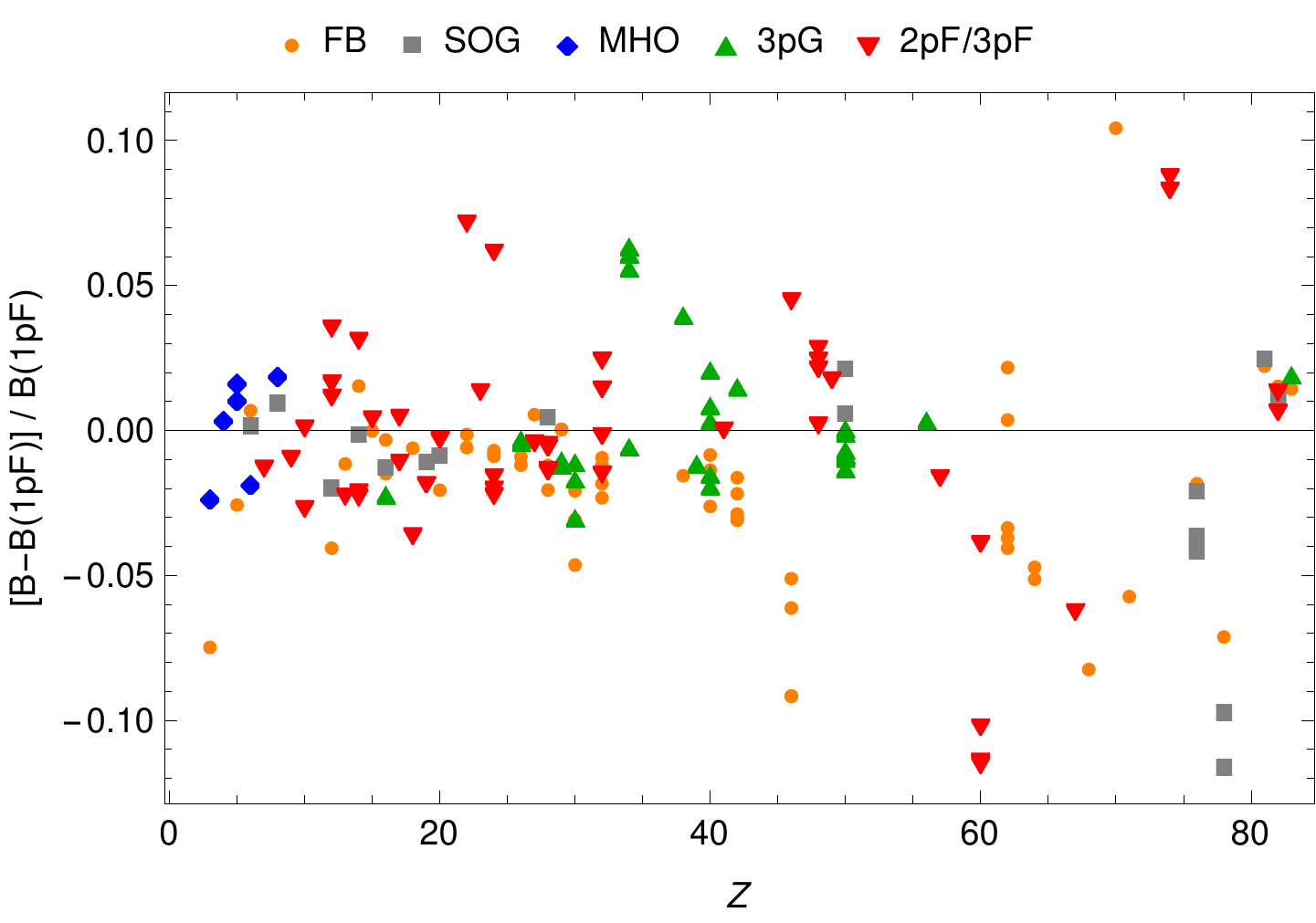}
\includegraphics[width=0.48\textwidth]{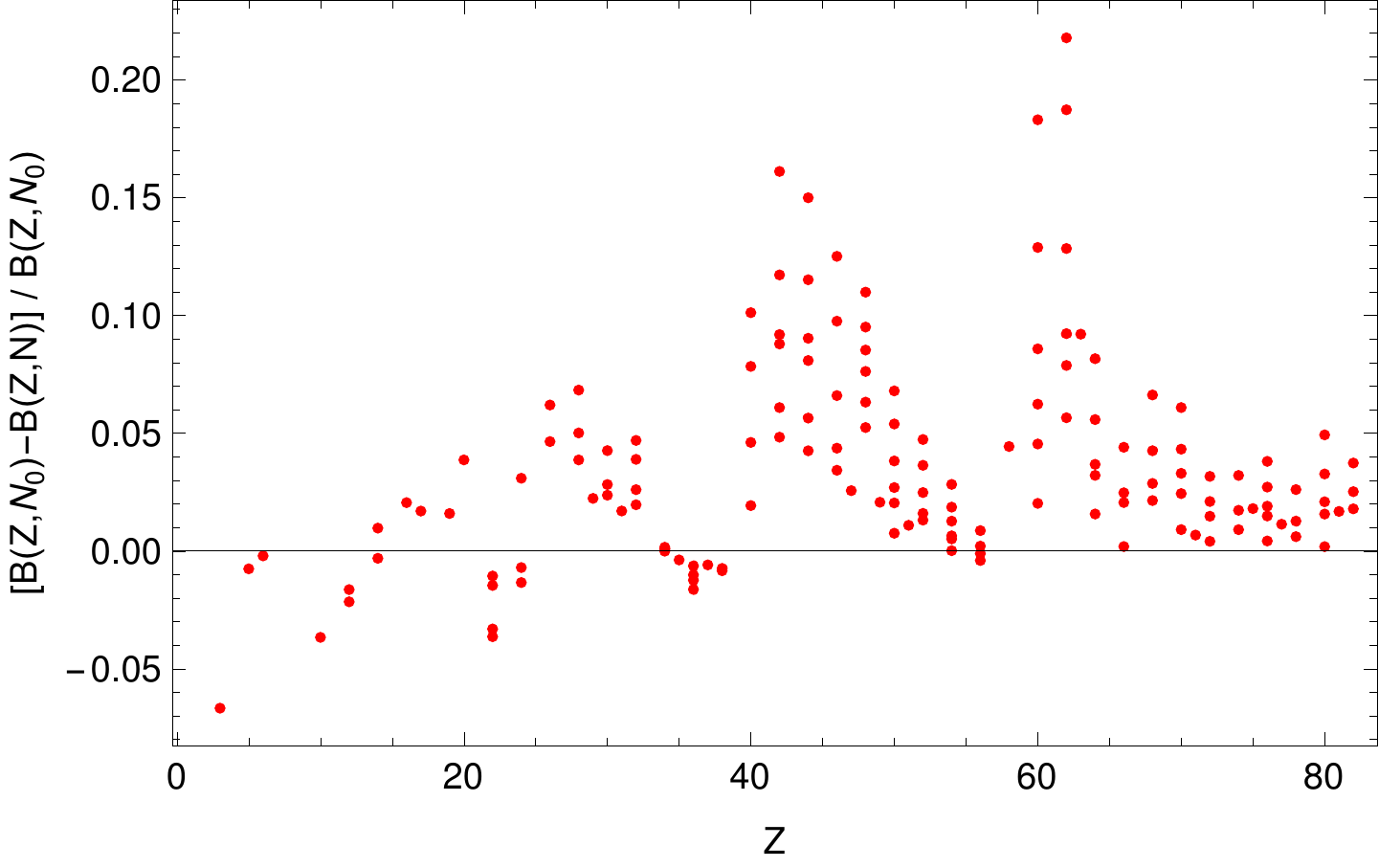}
\caption{
Top: relative difference of $B$ coefficients calculated with the 1pF charge distribution and other distributions.
Bottom: relative isotope dependence of $B$ calculated using 1pF distributions.
}
\label{fig:Bcoefficient_differences}
\end{figure}

To be more quantitative, we show in Fig.~\ref{fig:Bcoefficient_differences} (top) the differences in $B$ for charge parametrizations relative to 1pF. The differences for small $Z$ are typically below $5\%$ and grow to $10\%$ for larger $Z$. This serves as an estimate for the uncertainty on $B$ due to $\rho$.
In  Fig.~\ref{fig:Bcoefficient_differences} (bottom) we also see that the isotope differences are often of the same order or even larger than these uncertainties and should not be neglected.

\subsection{Connection to form factor}

\begin{figure}[t]
\includegraphics[width=0.48\textwidth]{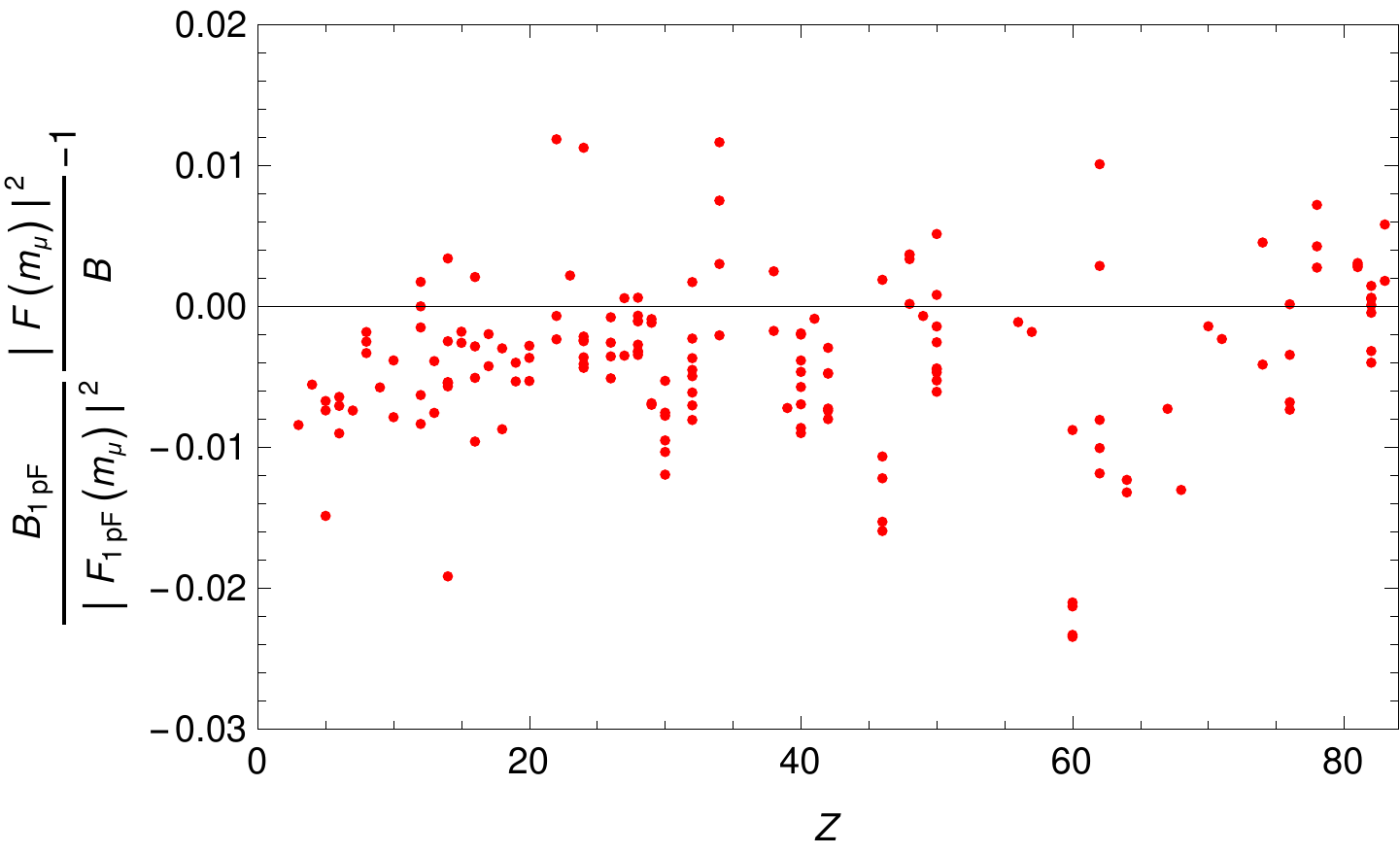}
\caption{
$B$ coefficient scaling with the squared form factor $|F(m_\mu)|^2$. See text for details.
}
\label{fig:formfactor}
\end{figure}

As mentioned above, the finite size of the nucleus in electron--nucleus scattering can, to first order, be incorporated by multiplying the point-nucleus cross section by the form factor squared at the relevant momentum transfer $q$: $|F(q)|^2$.
The same argument can be made for DIO, where we expect $B\propto |F(m_\mu)|^2$ at leading order, with $q= m_\mu$ the relevant momentum-transfer scale~\cite{Szafron:2015kja}.
Since the nuclear charge distribution complicates the numerical solution of the Dirac equation it would be convenient if this effect could indeed be factored out.
To test the scaling $B\propto |F(m_\mu)|^2$, we calculated the ratio of $B/|F(m_\mu)|^2$ for the 1pF charge distribution and compared it to the available multi-parameter charge distributions (FB, SOG, 3pG, 2pF/3pF). The result is shown in Fig.~\ref{fig:formfactor} and illustrates that the scaling $B\propto |F(m_\mu)|^2$ holds to better than $2\%$ for all $Z$, at least in the region of interest for us.
This makes it possible to simply rescale our $B$ results obtained with $\rho_\mathrm{1pF}$ by the ratio of form factors to obtain fairly accurate estimates of $B$ with new data, without the need to numerically solve Dirac equations again.

In addition, the scaling $B\propto |F(m_\mu)|^2$ opens up a different way to estimate the uncertainty on $B$ due to the charge distribution. Rather than trying to estimate the $B$ uncertainty from the uncertainty of $\rho(r)$ -- which is itself difficult to estimate as $\rho$ is not a measured observable -- we can look directly at electron--nucleus scattering data to obtain $|F(q=m_\mu)|^2$ including uncertainties. Despite the $\sim 2\%$ uncertainty in the scaling $B\propto |F(m_\mu)|^2$ (Fig.~\ref{fig:formfactor}) this can be beneficial compared to the usual chain of measuring $F(q)$, fitting $\rho(r)$, calculating $V(r)$, and solving Dirac equation to get $B$, which is quite tedious to propagate errors.
Notice that electron--nucleus scattering data around $q\sim m_\mu \simeq \unit[0.53]{fm^{-1}}$ is not available for all isotopes, nor is the actual form factor plus error bars given explicitly in many cases, rendering this approach unsuitable in general.

As a concrete example for the above procedure we consider \Element{Li}{3}{6}. The $B$ coefficients using 1pF and FB~\cite{Kabir:2015igz} are $B = \unit[1.20\times 10^{-19}]{MeV^{-6}}$ and $B = \unit[1.29\times 10^{-19}]{MeV^{-6}}$, respectively. Neither come with any reliable error bars so the best we can do is interpret the difference between the two parametrizations as an estimate of the uncertainty, which is roughly $7\%$.\footnote{Since \Element{Li}{3}{6} is a small nucleus ($r\simeq \unit[2.6]{fm}$), the 1pF parametrization with its constant skin thickness is not expected to be reliable here, so the FB result should be more realistic.}
Using instead the  form factor data from Ref.~\cite{Kabir:2015igz} underlying the FB parametrization, we have the interpolated value $|F(m_\mu)|^2 = 0.57$ with $2.5\%$ (systematic) uncertainty; since the experimental form factor value is larger than the FB fit at $q\sim m_\mu$~\cite{Kabir:2015igz}, $B$ should be larger as well, and can be estimated as
\begin{align}
B({}^6\text{Li}) &\simeq \frac{B_\text{FB}}{|F(m_\mu)|^2_\text{FB}}\, |F(m_\mu)|^2_\text{exp} \\
&= \frac{ \unit[1.29\times 10^{-19}]{MeV^{-6}}}{0.55}\,0.57 \\
&= \unit[1.32\times 10^{-19}]{MeV^{-6}}\,,
\end{align}
with an error of $3\%$, generously adding the experimental uncertainty and the $B\propto |F(m_\mu)|^2$ scaling uncertainty in quadrature.
This gives us a more reliable result for $B$ and its error, which in this case happens to be even smaller than our naive estimate.

\section{Conclusions}

The search for lepton flavor violation is one of our most sensitive probes of physics beyond the Standard Model. Experiments searching for $\mu^-\to e^-$ conversion such as COMET, DeeMe, and Mu2e, promise to improve existing limits by several orders of magnitude. An observation would make it possible to track the nucleus dependence of the $\mu^-\to e^-$ conversion rate by looking at different target materials, which would then help to discriminate the possible underlying new-physics models and operators.
Muon decay in orbit is an irreducible background in all $\mu^-\to e^-$ conversion searches and thus needs to be calculated to sufficient precision. In this article, we have performed a broad study of the DIO spectrum near the endpoint for all stable isotopes that could conceivably be used experimentally, extending previous studies. Our results are sufficient for background studies involving broad range of different or enriched target materials. 
Once the experiments choose alternative target materials, the next step in our investigation will be a detailed study of the DIO spectrum over the complete energy range.

\section*{Acknowledgements}
We thank all members of the Mu2e-II Snowmass21 Group for valuable discussions and support, especially Frank Porter, Lorenzo Calibbi, David Hitlin, Sophie Middleton, and Leo Borrell.
This work was partly supported by JSPS KAKENHI Grant Numbers JP18H01210 and JP21H00081 (Y.U.). R.S.~is supported by the United States Department of Energy under Grant Contract DE-SC0012704.

\appendix

\clearpage

\section{Analytic expressions for the potentials}
\label{app:potentials}

For completeness we give the normalization constants $\rho_0$ (defined via Eq.~\eqref{eq:normalization}) and the electric potentials $V$ (defined via Eq.~\eqref{eq:potential}) for all charge parametrizations of interest.

\subsection{Three-parameter Fermi model (3pF)}

The normalization is analytically given by
\begin{align}
\rho_0=& \frac{-Ze}{8\pi z^3\textrm{Li}_3\left(-\exp\frac{c}{z}\right)+\frac{96\pi\omega z^5}{c^2}\textrm{Li}_5\left(-\exp\frac{c}{z}\right)}
\end{align}
where $\mathrm{Li}_s$ is the $s$th-order polylogarithm function.
The analytic formula of the nuclear potential is
\begin{widetext}
\begin{align}
V(r)=& -e\rho_0\left[\frac{1}{r}\left\{\frac{c^3}{3}+\frac{\pi^2z^2c}{3}\right.\right. 
+\frac{\omega}{15c}\left(3c^4+10\pi^2z^2c^2+7\pi^4z^4\right)
-2z^3\left(\mathrm{Li}_3\left(-\chi_0\right)-\left(1+3\omega\frac{r^2}{c^2}\right)\mathrm{Li}_3\left(-\chi^{-1}\right)\right) \nonumber\\
&\hspace{1.5cm}\left.-24\omega\frac{z^5}{c^2}\left(\mathrm{Li}_5\left(-\chi_0\right)-\mathrm{Li}_5\left(-\chi^{-1}\right)\right)\right\} 
+z^2\left(1+\omega\frac{r^2}{c^2}\right)\mathrm{Li}_2\left(-\chi^{-1}\right) 
\left.+18\omega\frac{z^4}{c^2}\mathrm{Li}_4\left(-\chi^{-1}\right)\right],
\end{align}
where $\chi_0=\exp\left(-c/z\right)$ and $\chi=\chi_0\exp\left(r/z\right)$.
\end{widetext}

\subsection{Three-parameter Gaussian model (3pG)}

The normalization is
\begin{align}
\rho_0= \frac{-Ze/\left(\sqrt{\pi}z\right)^3}{\mathrm{Li}_{\frac{3}{2}}\left(-\exp\frac{c^2}{z^2}\right)+\omega\frac{3z^2}{2c^2}\mathrm{Li}_{\frac{5}{2}}\left(-\exp\frac{c^2}{z^2}\right)}\,.
\end{align}
The potential is
\begin{align}
V(r)=& -e\rho_0\left[I\left(r\right)+\frac{z^2}{2}\left(1+\omega\frac{r^2}{c^2}\right)\ln\left(1+\chi_G^{-1}\right)\right. \nonumber\\
&\hspace{1.5cm}\left.-\omega\frac{z^4}{2c^2}\mathrm{Li}_2\left(-\chi_G^{-1}\right)\right],
\end{align}
where $\chi_{G0}=\exp\left(-c^2/z^2\right)$ and $\chi_G=\chi_{G0}\exp\left(r^2/z^2\right)$.
Here the integral $I(r)$ is defined by
\begin{align}
I(r)=&\frac{z^3\chi_{G0}^{-1}}{r}\int_{0}^{r/z}\dd x\, x^2\frac{1+\omega\frac{x^2}{x_0^2}}{\exp\left(x^2\right)+\chi_{G0}^{-1}} 
\end{align}
with $x_0=c/z$.

\subsection{Modified-harmonic oscillator model (MHO)}

The normalization is
\begin{align}
\rho_0=&\frac{Ze}{4\pi}\frac{8}{a^3\sqrt{\pi}\left(2+3 w\right)}
\end{align}
and the potential takes the form
\begin{align}
V(r)=&-Z\alpha\left(\frac{\mathrm{erf}\left(\frac{r}{a}\right)}{r}-\frac{2w \exp\left(-\frac{r^2}{a^2}\right)}{a\sqrt{\pi}\left(2+3w\right)}\right),
\end{align}
where $\mathrm{erf}$ is the error function.

\subsection{Fourier--Bessel expansion (FB)}

The parameters $a_k$ ($1\le k \le n$) are normalized by
\begin{align}
\sum_{k}a_k\frac{\left(-1\right)^{k+1}}{k^2}=Ze\frac{\pi}{4R^3}.
\end{align}
The potential is
\begin{align}
V(r)=\begin{cases}
-\frac{Z\alpha}{R}-e\sum\limits_{k}a_k\frac{R^2}{k^2\pi^2}\frac{R}{k\pi r}\sin\frac{k\pi r}{R} \,, & r<R\,,\\
-\frac{Z\alpha}{r}\,, & r>R\,.
\end{cases}
\end{align}

\subsection{Sum of Gaussians (SOG)}

\begin{widetext}
The potential is
\begin{align}
V(r)=&-\frac{Z\alpha}{2r}\sum_{k}\frac{Q_k}{\gamma^2+2R_k^2} 
\left[\left\{\gamma^2-2R_k\left(r-R_k\right)\right\}\mathrm{erf}\left(\frac{r-R_k}{\gamma}\right)\right.
\left.+\left\{\gamma^2+2R_k\left(r+R_k\right)\right\}\mathrm{erf}\left(\frac{r+R_k}{\gamma}\right)\right. \nonumber\\
&+\frac{2\gamma R_k}{\sqrt{\pi}}\left\{\exp\left(-\frac{\left(r+R_k\right)^2}{\gamma^2}\right)\right.
\left.\left.-\exp\left(-\frac{\left(r-R_k\right)^2}{\gamma^2}\right)\right\}\right].
\end{align}
\end{widetext}

\pagebreak
\clearpage

\onecolumngrid
\section{Table of our results}
\label{app:table}

\bottomcaption{Table of our results, showing for each stable isotope with $Z>1$ and natural abundance above $1\%$ the endpoint energy $E_\text{end}'$ and $B$ coefficients as calculated using the various charge distributions.\label{tab:results}}
\tablehead{   & $E_\text{end}'/\unit{MeV}$ & $B_\text{1pF}/\unit{MeV^{-6}}$  & $B_\text{FB}/\unit{MeV^{-6}}$ & $B_\text{SOG}/\unit{MeV^{-6}}$ & $B_\text{MHO}/\unit{MeV^{-6}}$ & $B_\text{2pF/3pF}/\unit{MeV^{-6}}$& $B_\text{3pG}/\unit{MeV^{-6}}$\\ 
 \hline}
\begin{supertabular}{llllllll}
\Element{He}{2}{4} & $104.150$ &   &   & $2.53\times 10^{-20}$ &   &   &   \\
\Element{Li}{3}{6} & $104.637$ & $1.20\times 10^{-19}$ & $1.29\times 10^{-19}$ &   &   &   &   \\
\Element{Li}{3}{7} & $104.779$ & $1.28\times 10^{-19}$ &   &   & $1.31\times 10^{-19}$ &   &   \\
\Element{Be}{4}{9} & $104.949$ & $4.97\times 10^{-19}$ &   &   & $4.96\times 10^{-19}$ &   &   \\
\Element{B}{5}{10} & $104.99$ & $1.52\times 10^{-18}$ & $1.56\times 10^{-18}$ &   & $1.50\times 10^{-18}$ &   &   \\
\Element{B}{5}{11} & $105.044$ & $1.53\times 10^{-18}$ &   &   & $1.52\times 10^{-18}$ &   &   \\
\Element{C}{6}{12} & $105.059$ & $3.55\times 10^{-18}$ & $3.53\times 10^{-18}$ & $3.55\times 10^{-18}$ &   &   &   \\
\Element{C}{6}{13} & $105.097$ & $3.56\times 10^{-18}$ &   &   & $3.63\times 10^{-18}$ &   &   \\
\Element{N}{7}{14} & $105.094$ & $7.04\times 10^{-18}$ &   &   &   & $7.13\times 10^{-18}$ &   \\
\Element{O}{8}{16} & $105.106$ & $1.22\times 10^{-17}$ & $1.19\times 10^{-17}$ & $1.20\times 10^{-17}$ & $1.19\times 10^{-17}$ &   &   \\
\Element{F}{9}{19} & $105.118$ & $1.85\times 10^{-17}$ &   &   &   & $1.87\times 10^{-17}$ &   \\
\Element{Ne}{10}{20} & $105.081$ & $2.79\times 10^{-17}$ &   &   &   & $2.87\times 10^{-17}$ &   \\
\Element{Ne}{10}{22} & $105.108$ & $2.89\times 10^{-17}$ &   &   &   & $2.89\times 10^{-17}$ &   \\
\Element{Na}{11}{23} & $105.063$ & $4.35\times 10^{-17}$ &   &   &   &   &   \\
\Element{Mg}{12}{24} & $105.011$ & $6.17\times 10^{-17}$ &   & $6.29\times 10^{-17}$ &   & $6.10\times 10^{-17}$ &   \\
\Element{Mg}{12}{25} & $105.021$ & $6.30\times 10^{-17}$ &   &   &   & $6.08\times 10^{-17}$ &   \\
\Element{Mg}{12}{26} & $105.03$ & $6.27\times 10^{-17}$ & $6.52\times 10^{-17}$ &   &   & $6.17\times 10^{-17}$ &   \\
\Element{Al}{13}{27} & $104.971$ & $8.81\times 10^{-17}$ & $8.91\times 10^{-17}$ &   &   & $9.01\times 10^{-17}$ &   \\
\Element{Si}{14}{28} & $104.906$ & $1.17\times 10^{-16}$ & $1.19\times 10^{-16}$ & $1.17\times 10^{-16}$ &   & $1.20\times 10^{-16}$ &   \\
\Element{Si}{14}{29} & $104.914$ & $1.17\times 10^{-16}$ & $1.20\times 10^{-16}$ &   &   & $1.20\times 10^{-16}$ &   \\
\Element{Si}{14}{30} & $104.92$ & $1.16\times 10^{-16}$ & $1.14\times 10^{-16}$ &   &   & $1.12\times 10^{-16}$ &   \\
\Element{P}{15}{31} & $104.85$ & $1.50\times 10^{-16}$ & $1.50\times 10^{-16}$ &   &   & $1.50\times 10^{-16}$ &   \\
\Element{S}{16}{32} & $104.774$ & $1.88\times 10^{-16}$ & $1.90\times 10^{-16}$ & $1.90\times 10^{-16}$ &   &   & $1.92\times 10^{-16}$ \\
\Element{S}{16}{34} & $104.786$ & $1.84\times 10^{-16}$ & $1.84\times 10^{-16}$ &   &   &   &   \\
\Element{Cl}{17}{35} & $104.705$ & $2.23\times 10^{-16}$ &   &   &   & $2.22\times 10^{-16}$ &   \\
\Element{Cl}{17}{37} & $104.714$ & $2.19\times 10^{-16}$ &   &   &   & $2.21\times 10^{-16}$ &   \\
\Element{Ar}{18}{40} & $104.636$ & $2.69\times 10^{-16}$ & $2.70\times 10^{-16}$ &   &   & $2.79\times 10^{-16}$ &   \\
\Element{K}{19}{39} & $104.537$ & $3.36\times 10^{-16}$ &   & $3.40\times 10^{-16}$ &   & $3.42\times 10^{-16}$ &   \\
\Element{K}{19}{41} & $104.545$ & $3.31\times 10^{-16}$ &   &   &   &   &   \\
\Element{Ca}{20}{40} & $104.442$ & $4.01\times 10^{-16}$ & $4.09\times 10^{-16}$ & $4.04\times 10^{-16}$ &   & $4.02\times 10^{-16}$ &   \\
\Element{Ca}{20}{44} & $104.456$ & $3.85\times 10^{-16}$ &   &   &   &   &   \\
\Element{Sc}{21}{45} & $104.357$ & $4.59\times 10^{-16}$ &   &   &   &   &   \\
\Element{Ti}{22}{46} & $104.254$ & $5.24\times 10^{-16}$ &   &   &   &   &   \\
\Element{Ti}{22}{47} & $104.256$ & $5.30\times 10^{-16}$ &   &   &   &   &   \\
\Element{Ti}{22}{48} & $104.259$ & $5.32\times 10^{-16}$ & $5.35\times 10^{-16}$ &   &   & $4.94\times 10^{-16}$ &   \\
\Element{Ti}{22}{49} & $104.26$ & $5.41\times 10^{-16}$ &   &   &   &   &   \\
\Element{Ti}{22}{50} & $104.263$ & $5.43\times 10^{-16}$ & $5.44\times 10^{-16}$ &   &   &   &   \\
\Element{V}{23}{51} & $104.154$ & $6.33\times 10^{-16}$ &   &   &   & $6.24\times 10^{-16}$ &   \\
\Element{Cr}{24}{50} & $104.039$ & $7.09\times 10^{-16}$ & $7.14\times 10^{-16}$ &   &   & $7.23\times 10^{-16}$ &   \\
\Element{Cr}{24}{52} & $104.043$ & $7.18\times 10^{-16}$ & $7.25\times 10^{-16}$ &   &   & $7.35\times 10^{-16}$ &   \\
\Element{Cr}{24}{53} & $104.045$ & $7.14\times 10^{-16}$ &   &   &   & $6.70\times 10^{-16}$ &   \\
\Element{Cr}{24}{54} & $104.05$ & $6.87\times 10^{-16}$ & $6.92\times 10^{-16}$ &   &   & $6.98\times 10^{-16}$ &   \\
\Element{Mn}{25}{55} & $103.933$ & $7.95\times 10^{-16}$ &   &   &   &   &   \\
\Element{Fe}{26}{54} & $103.808$ & $9.42\times 10^{-16}$ & $9.51\times 10^{-16}$ &   &   &   & $9.45\times 10^{-16}$ \\
\Element{Fe}{26}{56} & $103.815$ & $8.98\times 10^{-16}$ & $9.09\times 10^{-16}$ &   &   &   & $9.03\times 10^{-16}$ \\
\Element{Fe}{26}{57} & $103.818$ & $8.84\times 10^{-16}$ &   &   &   &   &   \\
\Element{Co}{27}{59} & $103.698$ & $9.88\times 10^{-16}$ & $9.83\times 10^{-16}$ &   &   & $9.93\times 10^{-16}$ &   \\
\Element{Ni}{28}{58} & $103.566$ & $1.16\times 10^{-15}$ & $1.16\times 10^{-15}$ & $1.15\times 10^{-15}$ &   & $1.16\times 10^{-15}$ &   \\
\Element{Ni}{28}{60} & $103.572$ & $1.11\times 10^{-15}$ & $1.12\times 10^{-15}$ &   &   & $1.12\times 10^{-15}$ &   \\
\Element{Ni}{28}{61} & $103.575$ & $1.10\times 10^{-15}$ &   &   &   & $1.11\times 10^{-15}$ &   \\
\Element{Ni}{28}{62} & $103.578$ & $1.08\times 10^{-15}$ & $1.10\times 10^{-15}$ &   &   & $1.09\times 10^{-15}$ &   \\
\Element{Cu}{29}{63} & $103.451$ & $1.18\times 10^{-15}$ & $1.18\times 10^{-15}$ &   &   &   & $1.19\times 10^{-15}$ \\
\Element{Cu}{29}{65} & $103.456$ & $1.15\times 10^{-15}$ & $1.15\times 10^{-15}$ &   &   &   & $1.16\times 10^{-15}$ \\
\Element{Zn}{30}{64} & $103.322$ & $1.27\times 10^{-15}$ & $1.31\times 10^{-15}$ &   &   &   & $1.29\times 10^{-15}$ \\
\Element{Zn}{30}{66} & $103.327$ & $1.24\times 10^{-15}$ & $1.30\times 10^{-15}$ &   &   &   & $1.28\times 10^{-15}$ \\
\Element{Zn}{30}{67} & $103.328$ & $1.24\times 10^{-15}$ &   &   &   &   &   \\
\Element{Zn}{30}{68} & $103.331$ & $1.22\times 10^{-15}$ & $1.24\times 10^{-15}$ &   &   &   & $1.23\times 10^{-15}$ \\
\Element{Ga}{31}{69} & $103.198$ & $1.33\times 10^{-15}$ &   &   &   &   &   \\
\Element{Ga}{31}{71} & $103.202$ & $1.31\times 10^{-15}$ &   &   &   &   &   \\
\Element{Ge}{32}{70} & $103.064$ & $1.43\times 10^{-15}$ & $1.46\times 10^{-15}$ &   &   & $1.45\times 10^{-15}$ &   \\
\Element{Ge}{32}{72} & $103.069$ & $1.40\times 10^{-15}$ & $1.42\times 10^{-15}$ &   &   & $1.40\times 10^{-15}$ &   \\
\Element{Ge}{32}{73} & $103.07$ & $1.39\times 10^{-15}$ &   &   &   &   &   \\
\Element{Ge}{32}{74} & $103.073$ & $1.37\times 10^{-15}$ & $1.39\times 10^{-15}$ &   &   & $1.35\times 10^{-15}$ &   \\
\Element{Ge}{32}{76} & $103.076$ & $1.36\times 10^{-15}$ & $1.37\times 10^{-15}$ &   &   & $1.33\times 10^{-15}$ &   \\
\Element{As}{33}{75} & $102.934$ & $1.50\times 10^{-15}$ &   &   &   &   &   \\
\Element{Se}{34}{76} & $102.796$ & $1.59\times 10^{-15}$ &   &   &   &   & $1.50\times 10^{-15}$ \\
\Element{Se}{34}{77} & $102.797$ & $1.59\times 10^{-15}$ &   &   &   &   &   \\
\Element{Se}{34}{78} & $102.798$ & $1.58\times 10^{-15}$ &   &   &   &   & $1.49\times 10^{-15}$ \\
\Element{Se}{34}{80} & $102.8$ & $1.59\times 10^{-15}$ &   &   &   &   & $1.60\times 10^{-15}$ \\
\Element{Se}{34}{82} & $102.802$ & $1.59\times 10^{-15}$ &   &   &   &   & $1.49\times 10^{-15}$ \\
\Element{Br}{35}{79} & $102.655$ & $1.72\times 10^{-15}$ &   &   &   &   &   \\
\Element{Br}{35}{81} & $102.656$ & $1.72\times 10^{-15}$ &   &   &   &   &   \\
\Element{Kr}{36}{80} & $102.511$ & $1.83\times 10^{-15}$ &   &   &   &   &   \\
\Element{Kr}{36}{82} & $102.512$ & $1.84\times 10^{-15}$ &   &   &   &   &   \\
\Element{Kr}{36}{83} & $102.512$ & $1.85\times 10^{-15}$ &   &   &   &   &   \\
\Element{Kr}{36}{84} & $102.513$ & $1.84\times 10^{-15}$ &   &   &   &   &   \\
\Element{Kr}{36}{86} & $102.514$ & $1.85\times 10^{-15}$ &   &   &   &   &   \\
\Element{Rb}{37}{85} & $102.364$ & $2.00\times 10^{-15}$ &   &   &   &   &   \\
\Element{Rb}{37}{87} & $102.364$ & $2.01\times 10^{-15}$ &   &   &   &   &   \\
\Element{Sr}{38}{86} & $102.214$ & $2.13\times 10^{-15}$ &   &   &   &   &   \\
\Element{Sr}{38}{87} & $102.214$ & $2.14\times 10^{-15}$ &   &   &   &   &   \\
\Element{Sr}{38}{88} & $102.215$ & $2.15\times 10^{-15}$ & $2.18\times 10^{-15}$ &   &   &   & $2.06\times 10^{-15}$ \\
\Element{Y}{39}{89} & $102.062$ & $2.30\times 10^{-15}$ &   &   &   &   & $2.33\times 10^{-15}$ \\
\Element{Zr}{40}{90} & $101.909$ & $2.44\times 10^{-15}$ & $2.46\times 10^{-15}$ &   &   &   & $2.43\times 10^{-15}$ \\
\Element{Zr}{40}{91} & $101.913$ & $2.39\times 10^{-15}$ &   &   &   &   & $2.34\times 10^{-15}$ \\
\Element{Zr}{40}{92} & $101.918$ & $2.32\times 10^{-15}$ & $2.36\times 10^{-15}$ &   &   &   & $2.37\times 10^{-15}$ \\
\Element{Zr}{40}{94} & $101.926$ & $2.24\times 10^{-15}$ & $2.30\times 10^{-15}$ &   &   &   & $2.28\times 10^{-15}$ \\
\Element{Zr}{40}{96} & $101.932$ & $2.19\times 10^{-15}$ &   &   &   &   & $2.17\times 10^{-15}$ \\
\Element{Nb}{41}{93} & $101.762$ & $2.48\times 10^{-15}$ &   &   &   & $2.48\times 10^{-15}$ &   \\
\Element{Mo}{42}{92} & $101.595$ & $2.73\times 10^{-15}$ & $2.77\times 10^{-15}$ &   &   &   & $2.69\times 10^{-15}$ \\
\Element{Mo}{42}{94} & $101.607$ & $2.60\times 10^{-15}$ & $2.65\times 10^{-15}$ &   &   &   &   \\
\Element{Mo}{42}{95} & $101.61$ & $2.56\times 10^{-15}$ &   &   &   &   &   \\
\Element{Mo}{42}{96} & $101.616$ & $2.49\times 10^{-15}$ & $2.57\times 10^{-15}$ &   &   &   &   \\
\Element{Mo}{42}{97} & $101.618$ & $2.48\times 10^{-15}$ &   &   &   &   &   \\
\Element{Mo}{42}{98} & $101.624$ & $2.41\times 10^{-15}$ & $2.48\times 10^{-15}$ &   &   &   &   \\
\Element{Mo}{42}{100} & $101.635$ & $2.29\times 10^{-15}$ & $2.36\times 10^{-15}$ &   &   &   &   \\
\Element{Ru}{44}{96} & $101.286$ & $2.91\times 10^{-15}$ &   &   &   &   &   \\
\Element{Ru}{44}{98} & $101.296$ & $2.78\times 10^{-15}$ &   &   &   &   &   \\
\Element{Ru}{44}{99} & $101.3$ & $2.74\times 10^{-15}$ &   &   &   &   &   \\
\Element{Ru}{44}{100} & $101.306$ & $2.67\times 10^{-15}$ &   &   &   &   &   \\
\Element{Ru}{44}{101} & $101.309$ & $2.65\times 10^{-15}$ &   &   &   &   &   \\
\Element{Ru}{44}{102} & $101.316$ & $2.57\times 10^{-15}$ &   &   &   &   &   \\
\Element{Ru}{44}{104} & $101.325$ & $2.47\times 10^{-15}$ &   &   &   &   &   \\
\Element{Rh}{45}{103} & $101.153$ & $2.73\times 10^{-15}$ &   &   &   &   &   \\
\Element{Pd}{46}{102} & $100.978$ & $3.00\times 10^{-15}$ &   &   &   &   &   \\
\Element{Pd}{46}{104} & $100.988$ & $2.89\times 10^{-15}$ & $3.16\times 10^{-15}$ &   &   &   &   \\
\Element{Pd}{46}{105} & $100.99$ & $2.86\times 10^{-15}$ &   &   &   &   &   \\
\Element{Pd}{46}{106} & $100.996$ & $2.80\times 10^{-15}$ & $3.05\times 10^{-15}$ &   &   &   &   \\
\Element{Pd}{46}{108} & $101.005$ & $2.70\times 10^{-15}$ & $2.84\times 10^{-15}$ &   &   &   &   \\
\Element{Pd}{46}{110} & $101.013$ & $2.62\times 10^{-15}$ & $2.78\times 10^{-15}$ &   &   & $2.50\times 10^{-15}$ &   \\
\Element{Ag}{47}{107} & $100.83$ & $2.96\times 10^{-15}$ &   &   &   &   &   \\
\Element{Ag}{47}{109} & $100.838$ & $2.88\times 10^{-15}$ &   &   &   &   &   \\
\Element{Cd}{48}{106} & $100.654$ & $3.21\times 10^{-15}$ &   &   &   &   &   \\
\Element{Cd}{48}{110} & $100.67$ & $3.04\times 10^{-15}$ &   &   &   & $3.04\times 10^{-15}$ &   \\
\Element{Cd}{48}{111} & $100.673$ & $3.01\times 10^{-15}$ &   &   &   &   &   \\
\Element{Cd}{48}{112} & $100.677$ & $2.96\times 10^{-15}$ &   &   &   & $2.90\times 10^{-15}$ &   \\
\Element{Cd}{48}{113} & $100.68$ & $2.94\times 10^{-15}$ &   &   &   &   &   \\
\Element{Cd}{48}{114} & $100.683$ & $2.90\times 10^{-15}$ &   &   &   & $2.82\times 10^{-15}$ &   \\
\Element{Cd}{48}{116} & $100.688$ & $2.86\times 10^{-15}$ &   &   &   & $2.79\times 10^{-15}$ &   \\
\Element{In}{49}{113} & $100.506$ & $3.15\times 10^{-15}$ &   &   &   &   &   \\
\Element{In}{49}{115} & $100.513$ & $3.09\times 10^{-15}$ &   &   &   & $3.03\times 10^{-15}$ &   \\
\Element{Sn}{50}{116} & $100.341$ & $3.26\times 10^{-15}$ &   & $3.24\times 10^{-15}$ &   &   & $3.31\times 10^{-15}$ \\
\Element{Sn}{50}{117} & $100.344$ & $3.24\times 10^{-15}$ &   &   &   &   & $3.27\times 10^{-15}$ \\
\Element{Sn}{50}{118} & $100.348$ & $3.19\times 10^{-15}$ &   &   &   &   & $3.22\times 10^{-15}$ \\
\Element{Sn}{50}{119} & $100.35$ & $3.17\times 10^{-15}$ &   &   &   &   & $3.20\times 10^{-15}$ \\
\Element{Sn}{50}{120} & $100.353$ & $3.14\times 10^{-15}$ &   &   &   &   & $3.16\times 10^{-15}$ \\
\Element{Sn}{50}{122} & $100.359$ & $3.08\times 10^{-15}$ &   &   &   &   & $3.08\times 10^{-15}$ \\
\Element{Sn}{50}{124} & $100.363$ & $3.04\times 10^{-15}$ &   & $2.97\times 10^{-15}$ &   &   & $3.04\times 10^{-15}$ \\
\Element{Sb}{51}{121} & $100.189$ & $3.22\times 10^{-15}$ &   &   &   &   &   \\
\Element{Sb}{51}{123} & $100.193$ & $3.18\times 10^{-15}$ &   &   &   &   &   \\
\Element{Te}{52}{122} & $100.024$ & $3.29\times 10^{-15}$ &   &   &   &   &   \\
\Element{Te}{52}{124} & $100.028$ & $3.25\times 10^{-15}$ &   &   &   &   &   \\
\Element{Te}{52}{125} & $100.029$ & $3.24\times 10^{-15}$ &   &   &   &   &   \\
\Element{Te}{52}{126} & $100.032$ & $3.21\times 10^{-15}$ &   &   &   &   &   \\
\Element{Te}{52}{128} & $100.037$ & $3.17\times 10^{-15}$ &   &   &   &   &   \\
\Element{Te}{52}{130} & $100.04$ & $3.14\times 10^{-15}$ &   &   &   &   &   \\
\Element{I}{53}{127} & $99.8644$ & $3.30\times 10^{-15}$ &   &   &   &   &   \\
\Element{Xe}{54}{128} & $99.6974$ & $3.38\times 10^{-15}$ &   &   &   &   &   \\
\Element{Xe}{54}{129} & $99.6977$ & $3.38\times 10^{-15}$ &   &   &   &   &   \\
\Element{Xe}{54}{130} & $99.7001$ & $3.35\times 10^{-15}$ &   &   &   &   &   \\
\Element{Xe}{54}{131} & $99.6999$ & $3.36\times 10^{-15}$ &   &   &   &   &   \\
\Element{Xe}{54}{132} & $99.7026$ & $3.33\times 10^{-15}$ &   &   &   &   &   \\
\Element{Xe}{54}{134} & $99.7051$ & $3.31\times 10^{-15}$ &   &   &   &   &   \\
\Element{Xe}{54}{136} & $99.7087$ & $3.28\times 10^{-15}$ &   &   &   &   &   \\
\Element{Cs}{55}{133} & $99.5307$ & $3.45\times 10^{-15}$ &   &   &   &   &   \\
\Element{Ba}{56}{134} & $99.3628$ & $3.51\times 10^{-15}$ &   &   &   &   &   \\
\Element{Ba}{56}{135} & $99.3617$ & $3.52\times 10^{-15}$ &   &   &   &   &   \\
\Element{Ba}{56}{136} & $99.364$ & $3.50\times 10^{-15}$ &   &   &   &   &   \\
\Element{Ba}{56}{137} & $99.3633$ & $3.51\times 10^{-15}$ &   &   &   &   &   \\
\Element{Ba}{56}{138} & $99.3668$ & $3.48\times 10^{-15}$ &   &   &   &   & $3.47\times 10^{-15}$ \\
\Element{La}{57}{139} & $99.1928$ & $3.59\times 10^{-15}$ &   &   &   & $3.65\times 10^{-15}$ &   \\
\Element{Ce}{58}{140} & $99.0206$ & $3.68\times 10^{-15}$ &   &   &   &   &   \\
\Element{Ce}{58}{142} & $99.037$ & $3.51\times 10^{-15}$ &   &   &   &   &   \\
\Element{Pr}{59}{141} & $98.8438$ & $3.80\times 10^{-15}$ &   &   &   &   &   \\
\Element{Nd}{60}{142} & $98.6693$ & $3.89\times 10^{-15}$ &   &   &   & $4.04\times 10^{-15}$ &   \\
\Element{Nd}{60}{143} & $98.6772$ & $3.81\times 10^{-15}$ &   &   &   &   &   \\
\Element{Nd}{60}{144} & $98.6871$ & $3.71\times 10^{-15}$ &   &   &   & $4.14\times 10^{-15}$ &   \\
\Element{Nd}{60}{145} & $98.6939$ & $3.65\times 10^{-15}$ &   &   &   &   &   \\
\Element{Nd}{60}{146} & $98.7035$ & $3.56\times 10^{-15}$ &   &   &   & $3.92\times 10^{-15}$ &   \\
\Element{Nd}{60}{148} & $98.7213$ & $3.39\times 10^{-15}$ &   &   &   & $3.78\times 10^{-15}$ &   \\
\Element{Nd}{60}{150} & $98.7447$ & $3.18\times 10^{-15}$ &   &   &   & $3.55\times 10^{-15}$ &   \\
\Element{Sm}{62}{144} & $98.3182$ & $4.07\times 10^{-15}$ & $3.98\times 10^{-15}$ &   &   &   &   \\
\Element{Sm}{62}{147} & $98.3417$ & $3.84\times 10^{-15}$ &   &   &   &   &   \\
\Element{Sm}{62}{148} & $98.3512$ & $3.75\times 10^{-15}$ & $3.73\times 10^{-15}$ &   &   &   &   \\
\Element{Sm}{62}{149} & $98.3571$ & $3.69\times 10^{-15}$ &   &   &   &   &   \\
\Element{Sm}{62}{150} & $98.3728$ & $3.54\times 10^{-15}$ & $3.68\times 10^{-15}$ &   &   &   &   \\
\Element{Sm}{62}{152} & $98.3996$ & $3.31\times 10^{-15}$ & $3.44\times 10^{-15}$ &   &   &   &   \\
\Element{Sm}{62}{154} & $98.4142$ & $3.18\times 10^{-15}$ & $3.29\times 10^{-15}$ &   &   &   &   \\
\Element{Eu}{63}{151} & $98.194$ & $3.65\times 10^{-15}$ &   &   &   &   &   \\
\Element{Eu}{63}{153} & $98.2317$ & $3.32\times 10^{-15}$ &   &   &   &   &   \\
\Element{Gd}{64}{154} & $98.0516$ & $3.43\times 10^{-15}$ & $3.59\times 10^{-15}$ &   &   &   &   \\
\Element{Gd}{64}{155} & $98.058$ & $3.38\times 10^{-15}$ &   &   &   &   &   \\
\Element{Gd}{64}{156} & $98.0648$ & $3.32\times 10^{-15}$ &   &   &   &   &   \\
\Element{Gd}{64}{157} & $98.0668$ & $3.30\times 10^{-15}$ &   &   &   &   &   \\
\Element{Gd}{64}{158} & $98.0748$ & $3.24\times 10^{-15}$ & $3.40\times 10^{-15}$ &   &   &   &   \\
\Element{Gd}{64}{160} & $98.0858$ & $3.15\times 10^{-15}$ &   &   &   &   &   \\
\Element{Tb}{65}{159} & $97.8224$ & $3.99\times 10^{-15}$ &   &   &   &   &   \\
\Element{Dy}{66}{160} & $97.7249$ & $3.36\times 10^{-15}$ &   &   &   &   &   \\
\Element{Dy}{66}{161} & $97.7258$ & $3.35\times 10^{-15}$ &   &   &   &   &   \\
\Element{Dy}{66}{162} & $97.7337$ & $3.29\times 10^{-15}$ &   &   &   &   &   \\
\Element{Dy}{66}{163} & $97.7356$ & $3.27\times 10^{-15}$ &   &   &   &   &   \\
\Element{Dy}{66}{164} & $97.7438$ & $3.21\times 10^{-15}$ &   &   &   &   &   \\
\Element{Ho}{67}{165} & $97.5416$ & $3.48\times 10^{-15}$ &   &   &   & $3.70\times 10^{-15}$ &   \\
\Element{Er}{68}{164} & $97.3778$ & $3.43\times 10^{-15}$ &   &   &   &   &   \\
\Element{Er}{68}{166} & $97.3873$ & $3.36\times 10^{-15}$ & $3.63\times 10^{-15}$ &   &   &   &   \\
\Element{Er}{68}{167} & $97.3906$ & $3.33\times 10^{-15}$ &   &   &   &   &   \\
\Element{Er}{68}{168} & $97.3969$ & $3.28\times 10^{-15}$ &   &   &   &   &   \\
\Element{Er}{68}{170} & $97.4076$ & $3.20\times 10^{-15}$ &   &   &   &   &   \\
\Element{Tm}{69}{169} & $97.1782$ & $3.68\times 10^{-15}$ &   &   &   &   &   \\
\Element{Yb}{70}{170} & $97.032$ & $3.48\times 10^{-15}$ &   &   &   &   &   \\
\Element{Yb}{70}{171} & $97.0362$ & $3.45\times 10^{-15}$ &   &   &   &   &   \\
\Element{Yb}{70}{172} & $97.0431$ & $3.39\times 10^{-15}$ &   &   &   &   &   \\
\Element{Yb}{70}{173} & $97.0472$ & $3.36\times 10^{-15}$ &   &   &   &   &   \\
\Element{Yb}{70}{174} & $97.052$ & $3.33\times 10^{-15}$ & $2.98\times 10^{-15}$ &   &   &   &   \\
\Element{Yb}{70}{176} & $97.0604$ & $3.27\times 10^{-15}$ &   &   &   &   &   \\
\Element{Lu}{71}{175} & $96.9069$ & $3.14\times 10^{-15}$ & $3.32\times 10^{-15}$ &   &   &   &   \\
\Element{Lu}{71}{176} & $96.9101$ & $3.12\times 10^{-15}$ &   &   &   &   &   \\
\Element{Hf}{72}{176} & $96.6831$ & $3.53\times 10^{-15}$ &   &   &   &   &   \\
\Element{Hf}{72}{177} & $96.6852$ & $3.51\times 10^{-15}$ &   &   &   &   &   \\
\Element{Hf}{72}{178} & $96.6903$ & $3.48\times 10^{-15}$ &   &   &   &   &   \\
\Element{Hf}{72}{179} & $96.6934$ & $3.46\times 10^{-15}$ &   &   &   &   &   \\
\Element{Hf}{72}{180} & $96.6985$ & $3.42\times 10^{-15}$ &   &   &   &   &   \\
\Element{Ta}{73}{181} & $96.5092$ & $3.55\times 10^{-15}$ &   &   &   &   &   \\
\Element{W}{74}{182} & $96.3202$ & $3.67\times 10^{-15}$ &   &   &   &   &   \\
\Element{W}{74}{183} & $96.3248$ & $3.64\times 10^{-15}$ &   &   &   &   &   \\
\Element{W}{74}{184} & $96.3289$ & $3.61\times 10^{-15}$ &   &   &   & $3.31\times 10^{-15}$ &   \\
\Element{W}{74}{186} & $96.3363$ & $3.55\times 10^{-15}$ &   &   &   & $3.24\times 10^{-15}$ &   \\
\Element{Re}{75}{185} & $96.1295$ & $3.81\times 10^{-15}$ &   &   &   &   &   \\
\Element{Re}{75}{187} & $96.1386$ & $3.74\times 10^{-15}$ &   &   &   &   &   \\
\Element{Os}{76}{186} & $95.9621$ & $3.76\times 10^{-15}$ &   &   &   &   &   \\
\Element{Os}{76}{187} & $95.9644$ & $3.74\times 10^{-15}$ &   &   &   &   &   \\
\Element{Os}{76}{188} & $95.9698$ & $3.70\times 10^{-15}$ &   & $3.86\times 10^{-15}$ &   &   &   \\
\Element{Os}{76}{189} & $95.972$ & $3.69\times 10^{-15}$ &   &   &   &   &   \\
\Element{Os}{76}{190} & $95.9762$ & $3.66\times 10^{-15}$ &   & $3.79\times 10^{-15}$ &   &   &   \\
\Element{Os}{76}{192} & $95.9821$ & $3.62\times 10^{-15}$ & $3.68\times 10^{-15}$ & $3.69\times 10^{-15}$ &   &   &   \\
\Element{Ir}{77}{191} & $95.7727$ & $3.88\times 10^{-15}$ &   &   &   &   &   \\
\Element{Ir}{77}{193} & $95.7787$ & $3.83\times 10^{-15}$ &   &   &   &   &   \\
\Element{Pt}{78}{194} & $95.6016$ & $3.85\times 10^{-15}$ &   & $4.23\times 10^{-15}$ &   &   &   \\
\Element{Pt}{78}{195} & $95.6048$ & $3.83\times 10^{-15}$ &   &   &   &   &   \\
\Element{Pt}{78}{196} & $95.6084$ & $3.80\times 10^{-15}$ & $4.07\times 10^{-15}$ & $4.24\times 10^{-15}$ &   &   &   \\
\Element{Pt}{78}{198} & $95.6157$ & $3.75\times 10^{-15}$ &   &   &   &   &   \\
\Element{Au}{79}{197} & $95.4181$ & $3.91\times 10^{-15}$ &   &   &   &   &   \\
\Element{Hg}{80}{198} & $95.2299$ & $4.01\times 10^{-15}$ &   &   &   &   &   \\
\Element{Hg}{80}{199} & $95.2311$ & $4.00\times 10^{-15}$ &   &   &   &   &   \\
\Element{Hg}{80}{200} & $95.2387$ & $3.94\times 10^{-15}$ &   &   &   &   &   \\
\Element{Hg}{80}{201} & $95.2417$ & $3.92\times 10^{-15}$ &   &   &   &   &   \\
\Element{Hg}{80}{202} & $95.2483$ & $3.87\times 10^{-15}$ &   &   &   &   &   \\
\Element{Hg}{80}{204} & $95.2578$ & $3.81\times 10^{-15}$ &   &   &   &   &   \\
\Element{Tl}{81}{203} & $95.0527$ & $4.02\times 10^{-15}$ & $3.92\times 10^{-15}$ &   &   &   &   \\
\Element{Tl}{81}{205} & $95.0621$ & $3.95\times 10^{-15}$ & $3.86\times 10^{-15}$ & $3.85\times 10^{-15}$ &   &   &   \\
\Element{Pb}{82}{204} & $94.8683$ & $4.07\times 10^{-15}$ & $4.03\times 10^{-15}$ &   &   &   &   \\
\Element{Pb}{82}{206} & $94.8786$ & $4.00\times 10^{-15}$ & $3.94\times 10^{-15}$ & $3.96\times 10^{-15}$ &   & $3.98\times 10^{-15}$ &   \\
\Element{Pb}{82}{207} & $94.8828$ & $3.97\times 10^{-15}$ & $3.92\times 10^{-15}$ &   &   & $3.92\times 10^{-15}$ &   \\
\Element{Pb}{82}{208} & $94.8899$ & $3.92\times 10^{-15}$ & $3.88\times 10^{-15}$ & $3.87\times 10^{-15}$ &   &   &   \\
\Element{Bi}{83}{209} & $94.7121$ & $3.93\times 10^{-15}$ & $3.87\times 10^{-15}$ &   &   &   & $3.86\times 10^{-15}$ \\
\Element{Th}{90}{232} & $93.6143$ & $3.07\times 10^{-15}$ &   &   &   &   &   \\
\Element{U}{92}{238} & $93.3036$ & $2.85\times 10^{-15}$ &   &   &   & $4.07\times 10^{-15}$ &   \\
\hline
\end{supertabular}

\newpage
\bibliographystyle{utcaps_mod}
\bibliography{BIB}

\end{document}